\documentclass[10pt]{article}
\usepackage{amsmath,amsfonts}
\usepackage{graphicx}
\usepackage{epsfig,epsf}
\usepackage{hyperref}
\usepackage{bm}

\def\l {\lambda}

\def\bar {\overline}

\def\be {\begin{equation}}
\def\ee {\end{equation}}
\def\beq {\begin{equation}}
\def\eeq {\end{equation}}
\def\bea {\begin{eqnarray}}
\def\eea {\end{eqnarray}}

\def\beq{\begin{equation}}
\def\eeq{\end{equation}}
\def\barr{\begin{array}}
\def\earr{\end{array}}

\def\opcit(#1){ {\em op. cit.}, #1}

\def\issue(#1,#2,#3){#1, #2 (#3)} 

\def\APP(#1,#2,#3){Acta Phys.\ Polon.\ \issue(#1,#2,#3)}
\def\ARNPS(#1,#2,#3){Ann.\ Rev.\ Nucl.\ Part.\ Sci.\ \issue(#1,#2,#3)}
\def\CPC(#1,#2,#3){Comp.\ Phys.\ Comm.\ \issue(#1,#2,#3)}
\def\CIP(#1,#2,#3){Comput.\ Phys.\ \issue(#1,#2,#3)}
\def\EPJC(#1,#2,#3){Eur.\ Phys.\ J.\ C\ \issue(#1,#2,#3)}
\def\EPJD(#1,#2,#3){Eur.\ Phys.\ J. Direct\ C\ \issue(#1,#2,#3)}
\def\IEEETNS(#1,#2,#3){IEEE Trans.\ Nucl.\ Sci.\ \issue(#1,#2,#3)}
\def\IJMP(#1,#2,#3){Int.\ J.\ Mod.\ Phys. \issue(#1,#2,#3)}
\def\JHEP(#1,#2,#3){J.\ High Energy Physics \issue(#1,#2,#3)}
\def\JPG(#1,#2,#3){J.\ Phys.\ G \issue(#1,#2,#3)}
\def\MPL(#1,#2,#3){Mod.\ Phys.\ Lett.\ \issue(#1,#2,#3)}
\def\NP(#1,#2,#3){Nucl.\ Phys.\ \issue(#1,#2,#3)}
\def\NIM(#1,#2,#3){Nucl.\ Instrum.\ Meth.\ \issue(#1,#2,#3)}
\def\PL(#1,#2,#3){Phys.\ Lett.\ \issue(#1,#2,#3)}
\def\PRD(#1,#2,#3){Phys.\ Rev.\ D \issue(#1,#2,#3)}
\def\PRL(#1,#2,#3){Phys.\ Rev.\ Lett.\ \issue(#1,#2,#3)}
\def\SJNP(#1,#2,#3){Sov.\ J. Nucl.\ Phys.\ \issue(#1,#2,#3)}
\def\ZPC(#1,#2,#3){Zeit.\ Phys.\ C \issue(#1,#2,#3)}

\begin{document}

\renewcommand*{\thefootnote}{\fnsymbol{footnote}}

\begin{center}
 {\Large\bf{Diphoton excess at 750 GeV:
 Singlet scalars confront triviality}}

\vspace{3mm}

Indrani Chakraborty \footnote{indrani300888@gmail.com} and 
Anirban Kundu \footnote{anirban.kundu.cu@gmail.com}

\vspace{1mm}
{\em{Department of Physics, University of Calcutta, \\
92 Acharya Prafulla Chandra Road, Kolkata 700009, India
}}

\end{center}
\begin{abstract}
 We suggest that the recently observed diphoton excess at 750 GeV comes from a 
 quasi-degenerate bunch of gauge singlet scalars produced and decaying through one 
 or more vector-like fermions. This explains the broad nature of the resonance, 
 even though the decay is loop-mediated. At the same time, the model keeps the new 
 Yukawa couplings in the perturbative region, which is necessary for the stability 
 of the potential. 

\end{abstract}

\date{\today}



\setcounter{footnote}{0}
\renewcommand*{\thefootnote}{\arabic{footnote}}


\section{Introduction}

Recently, the ATLAS and the CMS Collaborations at the Large Hadron Collider (LHC) 
announced the hint of a resonance seen at 
around $\sqrt{s}=750$ GeV, and decaying into two photons \cite{atlas-750,cms-750}. The local significance 
of this event is $3.6\sigma$ for ATLAS, and $2.6\sigma$ for CMS; the global significance, taking into account 
the look-elsewhere effect, is somewhat smaller.
The combined local significance of such a diphoton excess is above $4\sigma$. While it is too early 
to definitely predict any new physics beyond the Standard Model (SM) right now, the result 
has led to a flurry of different interpretations. The two-photon decay channel excludes spin-1 
nature of the resonance, leaving open spin-0 or spin-2 options (theoretically, higher spins too). 
Even these options are highly constrained from the apparent significance of the signal even with 
such a low integrated luminosity, which points to rather strong couplings, and also from 
non-observation of excesses in certain channels, like dilepton or $t\bar{t}$. 

In this paper, we will focus on one of the simplest and most economical explanations of the 
excess, that caused by a gauge singlet scalar (or a bunch of such scalars, degenerate in mass). 
For production through gluon fusion and subsequent 
decay to two photons, this new particle must couple with some vectorlike fermions (VF). Such a model 
has been discussed in Refs.\ \cite{bhaskar,falkowski,singlet750}; there have been other proposals 
using one or more VFs \cite{VF}. Examples of models with such vectorlike fermions can be found, {\em e.g.}, 
in Refs.\ \cite{VFpheno}. 

While being economical, the model 
also raises a few questions. First, the new Yukawa couplings of the VFs 
with the scalar have to be very large to explain the signal. Such couplings are not only 
nonperturbative but also make the scalar potential unstable because of the strong negative pull 
on the singlet quartic coupling, which must be positive for stability of the potential. At the same time,
it is hard to predict why the width of such a loop-mediated decay be broad. One must mention that 
attempts to explain this signal are aplenty in the literature \cite{goldrush} \footnote{ 
This is reminiscent of the initial Higgs boson results, where the diphoton decay width was slightly 
anomalous \cite{carmi,jaeckel}, and which gave rise to a lot of possible new physics interpretations.}. 
Even when one takes the weighted average of 8 TeV and 13 TeV data, the Yukawa couplings remain 
nonperturbative. 

We would like to ask the question: assuming that the model is renormalizable, how does one make the 
scalar potential stable? In other words, how does one make the Yukawa couplings small and still be 
consistent with the data? One answer, of course, is to introduce more VFs; the number of VFs depend on 
their quantum numbers as well as what is considered to be the perturbative limit. The production cross-section 
goes as the square of the VF Yukawa coupling, while the negative pull on the scalar quartic coupling 
goes as the fourth power of the same (the expressions are shown later), so obviously introduction of more VFs 
will help. An alternative option, which we would like to explore, is to introduce more than one 
gauge singlet scalars. While such a plethora of scalars may appear bizarre at the first sight, it is 
perhaps not more bizarre than an equally imposing plethora of VFs.

The simplest version of such a model with $N$ singlet scalars 
may have a degeneracy or quasi-degeneracy among them. This may also be motivated by some $O(N)$ 
symmetry of the potential, whose soft breaking possibly gives rise to a quasi-degeneracy of mass. 
If these scalars are quasi-degenerate, it explains why the resonance looks broad; this is actually a bunch 
of closely-spaced narrow resonances, seen through experiments whose energy resolution is not that fine. 
Such a fake broad resonance is hard to resolve at the LHC unless there are two distinct bumps at least 20 GeV 
apart. Multiple resonances 
also mean that the Yukawa couplings can be small, the effect being the sum of all these resonances. This 
makes the model stable with respect to a renormalization group running until the couplings blow up at the 
Landau pole. 

In Section II, we briefly discuss the model, and show our results in Section III. In Section IV, we 
summarize and conclude.

\section{The model}

As this is a short paper, we will not go through the salient features of the model but refer the reader 
to Ref.\ \cite{ic1} for details. The scalar potential is
\be
V(\Phi,S_i) = -\mu^2\Phi^\dag\Phi + \lambda (\Phi^\dag\Phi)^2 
+M^2 \sum_i S_i^2 + \lambda_S \left(\sum_i S_i^2\right)^2 + a (\Phi^\dag\Phi)\sum_i S_i^2\,.
\label{multising}
\ee
where, apart from the SM doublet $\Phi$, we have taken $N$ number of gauge singlet scalars, $i=1 \cdots N$. 
There is an explicit $Z_2$ symmetry, $S_i \to -S_i$, preventing the odd terms in $S_i$. We will take  
$\mu^2, M^2 > 0$ so that the $S$-fields do not have a nonzero vacuum expectation value (VEV) and hence 
do not mix with $\Phi$. At this simplest level, all singlets are degenerate and have the same quartic 
coupling ${\l}_S$. This, however, is not a strict requirement. One can also remove the ad hoc $Z_2$ symmetry.
That will allow terms like $\Phi^\dag\Phi S_i$ in the potential and may lead to $S_i$ decaying into two Higgs bosons. 
Without the $Z_2$ symmetry, the potential is also far richer and may involve multiple minima of different depth. 

For the time being, we will consider only one vector quark $Q$, singlet under 
$SU(2)_L$ and with electric charge $e_Q$ , which can 
be $+\frac23$ or $-\frac13$ for conventional vector fermion models. This brings in two further terms in the 
potential:
\be
{\cal L}_Q \supset -M_Q \bar{Q}Q - \zeta_{iQ} \bar{Q} Q S_i\,,
\ee
where $\zeta_Q$, the new Yukawa coupling, plays the crucial role in production and decay of the singlets. For 
simplicity, we take all the $\zeta_{iQ}$s to be the same and denote it by $\zeta_Q$. 

Knowledge of the potential parameters like $\lambda_S$, $a$, or $\zeta_Q$ will help us narrowing down the 
features of the potential \cite{sg}. However, only $\zeta_Q$ can be accessed through the diphoton channel and 
the quartic couplings $a$ and $\lambda_S$ remain free parameters of the theory. 

There are no constraints 
from oblique parameters if the VFs are degenerate and the singlet scalars do not mix with $\Phi$. 
The stability of the potential imposes the following conditions for stability:
\be
\lambda > 0\,,\ \ \ \lambda_S > 0\,,\ \ \ a+2\sqrt{\lambda\lambda_S} > 0\,.
\ee

Next, we would like to see how the couplings evolve with energy. 
We will limit our discussions 
within one-loop effects only
The one-loop $\beta$-functions are 
\cite{ic1}
\bea
16\pi^2 \beta_\lambda &=& 12\lambda^2 + 6g_t^2\lambda + N a^2 -\frac32\lambda(g_1^2+3g_2^2) 
- 3 g_t^4 + \frac{3}{16}(g_1^4 + 2 g_1^2 g_2^2 + 3 g_2^4)\,,\nonumber\\
16\pi^2 \beta_{\lambda_S} &=& (32 + 4N) {\lambda_S}^2 + a^2 + 4\lambda_S Z^2 - N_c \zeta_Q^4
\,,\nonumber\\
16\pi^2 \beta_a &=& \left[ 6\lambda + 12\lambda_S + 4a + 6g_t^2 + 4Z^2 -\frac32 g_1^2 -\frac92 g_2^2\right]a
\,,\nonumber\\
16\pi^2 \beta_{g_t} &=& \left[ \frac94 g_t^2 - \frac{17}{24}g_1^2 - \frac98 g_2^2 - 4g_3^2\right] g_t\,,
\nonumber\\
16\pi^2\beta_{g_3} &=& -\frac{17}{6} g_3^3 \theta(q^2-m_Q^2) - \frac{19}{6} g_3^3 \theta(m_Q^2-q^2)\,,\nonumber\\
16\pi^2 \beta_{\zeta_U} &=& \left[ \frac32 \zeta_U^2 + Z^2 -\frac43\left(\frac{1}{12}\right) g_1^2 - 0
\left(\frac{9}{4}\right) g_2^2 - 4g_3^2\right]\zeta_U\,,\nonumber\\
16\pi^2 \beta_{\zeta_D} &=& \left[ \frac32 \zeta_U^2 + Z^2 -\frac13\left(\frac{1}{12}\right) g_1^2 - 0
\left(\frac{9}{4}\right) g_2^2 - 4g_3^2\right]\zeta_D\,,\nonumber\\
\label{all-rge}
\eea
where $\beta_h\equiv dh/dt$, and $t \equiv \ln(q^2/\mu^2)$, and we have taken all VFs to be 
heavier than the top. If there are more than one such VFs, the last term in the second $\beta$-function, 
$N_c\zeta_Q^4$, should be replaced by $\sum_i N_c^i \zeta_i^4$, where $N_c^i=3(1)$ for quarks (leptons).
Note that our definition of $t$ differs 
by a factor of 2 from that used by some authors. For the new fermions, the $\beta$-functions 
are given for the singlet (doublet) type VFs. For simplicity, we have put all the SM Yukawa 
couplings equal to zero except for that of the top quark. This hardly changes our conclusions. Note that 
a large value of $\zeta_Q$ quickly makes $\lambda_S$ negative because of the $\zeta_Q^4$ term, thus rendering 
the potential unstable.

If the vectorlike quark $Q$ is much heavier than 750 GeV, we can integrate it out to write the 
effective interaction vertices 
\be
{\cal L}_{\rm eff} = C_\gamma F_{\mu\nu}F^{\mu\nu} S + C_g G^a_{\mu\nu} G^{a\mu\nu} S\,,
\ee
where
\bea
C_\gamma &=& \frac{\alpha}{2\pi} N_c e_Q^2 \frac{\zeta_Q}{M_Q} A_{1/2}(x)\,,\nonumber\\
C_g &=& \frac{\alpha_s}{4\pi} \frac{\zeta_Q}{M_Q} A_{1/2}(x)\,,
\eea
with
\be
A_{1/2}(x) = 2x\left[ 1+(1-x)f(x)\right]\,,\ \ 
f(x) = \left[ \sin^{-1}(1/\sqrt{x})\right]^2\,.
\ee
Here $x=4M_Q^2/M_S^2$ and in the limit $x \gg 1$, $A_{1/2}(x) \to \frac43$.  
The decay widths in this limit are given by 
\be
\Gamma(S\to\gamma\gamma) = \frac{\alpha^2}{16\pi^3} e_Q^4 \zeta_Q^2 \frac{M_S^3}{M_Q^2}\,,\ \ 
\Gamma(S\to gg) = \frac{\alpha_s^2}{72\pi^3} \zeta_Q^2 \frac{M_S^3}{M_Q^2}\,.
\ee
The cross-section for $pp\to \gamma\gamma$, 
mediated by all the scalars, is proportional to
\be
\sigma(pp\to\gamma\gamma) \propto \frac{N}{M_S} \left( \frac{\Gamma(S\to \gamma\gamma) \Gamma(S\to gg)}
{\Gamma(S\to \gamma\gamma) + \Gamma(S\to gg)} \right) \,.
\ee

\section{Analysis}

We will not go into a detailed analysis of the signal here. The 13 TeV data, in a narrow-width approximation, 
gives $\sigma(pp\to\gamma\gamma) \in [3:9]$ fb at 68\% CL, taking both ATLAS and CMS numbers 
\cite{falkowski}, for $M_S=750$ GeV. This gives a range of $\zeta_Q$, as a function of the number of singlets
$N$, the mass of the fermion, and the electric charge of the fermion. Taking both 8 TeV and 13 TeV data, 
the range for the cross-section, at 68\% CL, changes to $[1.3:4.2]$ fb. We will work only with the 13 TeV data. 

\begin{figure}[t!]
\begin{center} 
\includegraphics[width=8.5cm]{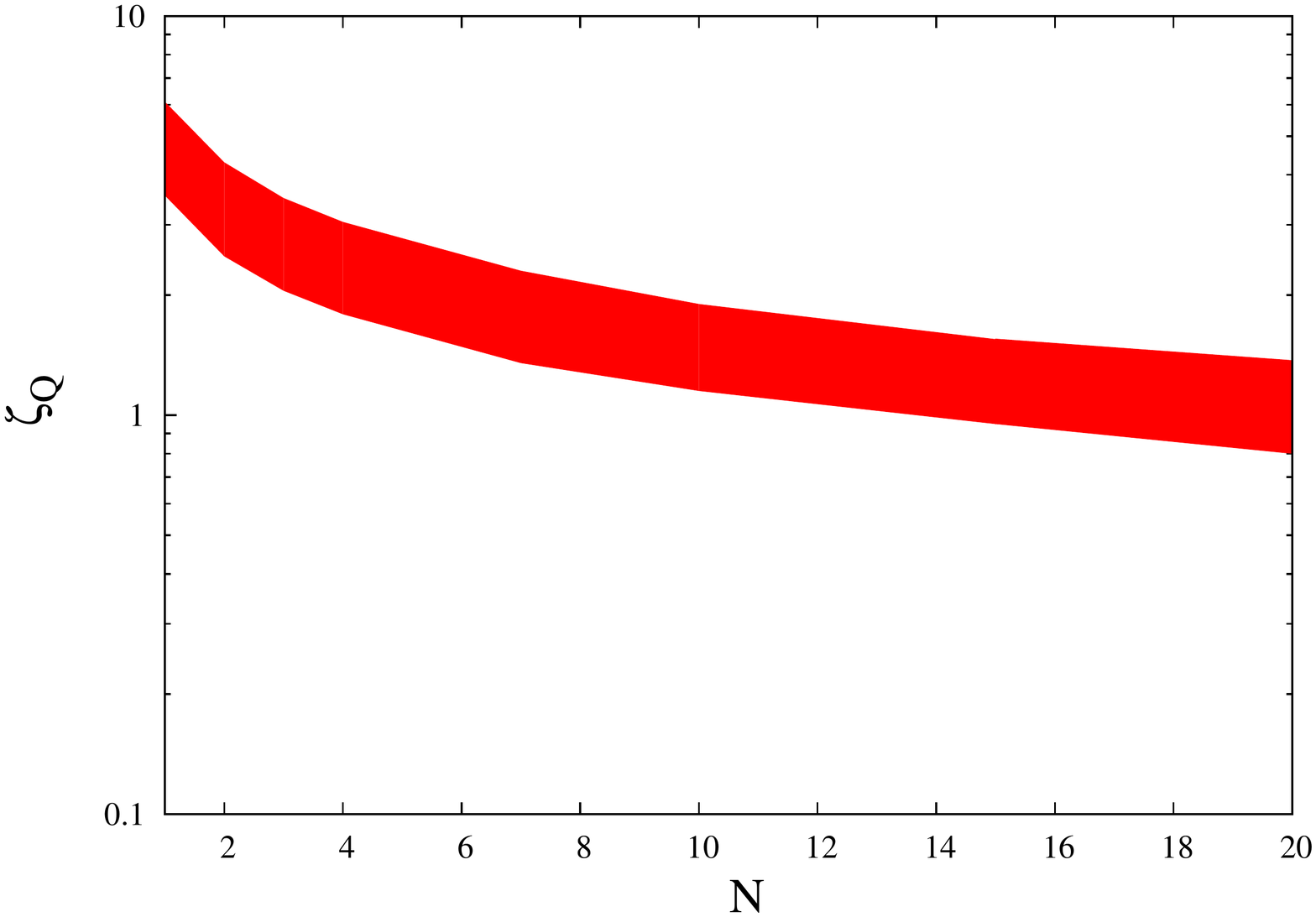}
\end{center}
\caption{\em\small Upper and lower ranges of the Yukawa coupling $\zeta_Q$, as a function of the number of 
singlet scalars $N$, assuming a diphoton cross section 
of $[3:9]$ fb. }
\label{fig:zeta}
\end{figure}

In Fig.\ \ref{fig:zeta}, we show the range of the Yukawa coupling $\zeta_Q$ that is needed to generate the signal, 
as a function of the number of singlet scalars. The range is highly nonperturbative for a small number of scalars. 
The plot is drawn for $M_Q = 1$ TeV and $e_Q = +\frac23$, it gets even worse for heavier fermions (only the ratio
$\zeta_Q/M_Q$ is bounded from the data), and down-type quarks, for which the branching fraction to $\gamma\gamma$ 
goes down, needing an even higher $\zeta_Q$ to compensate it. One may add a full vectorial generation, including 
leptons, to add a further channel that helps in $S\to\gamma\gamma$ decay but does not affect the production
\cite{bhaskar}, but a single VF generation with one singlet scalar still needs uncomfortably large $\zeta_Q$. 
This is only to be expected, as we are trying to reproduce a signal, whose strength is more indicative 
of a strong dynamics, through weak dynamics only. With a single scalar, it is also impossible to reproduce a large decay
width with perturbative Yukawa couplings. 

\begin{figure}[t!]
\begin{center} 
\includegraphics[width=5.5cm]{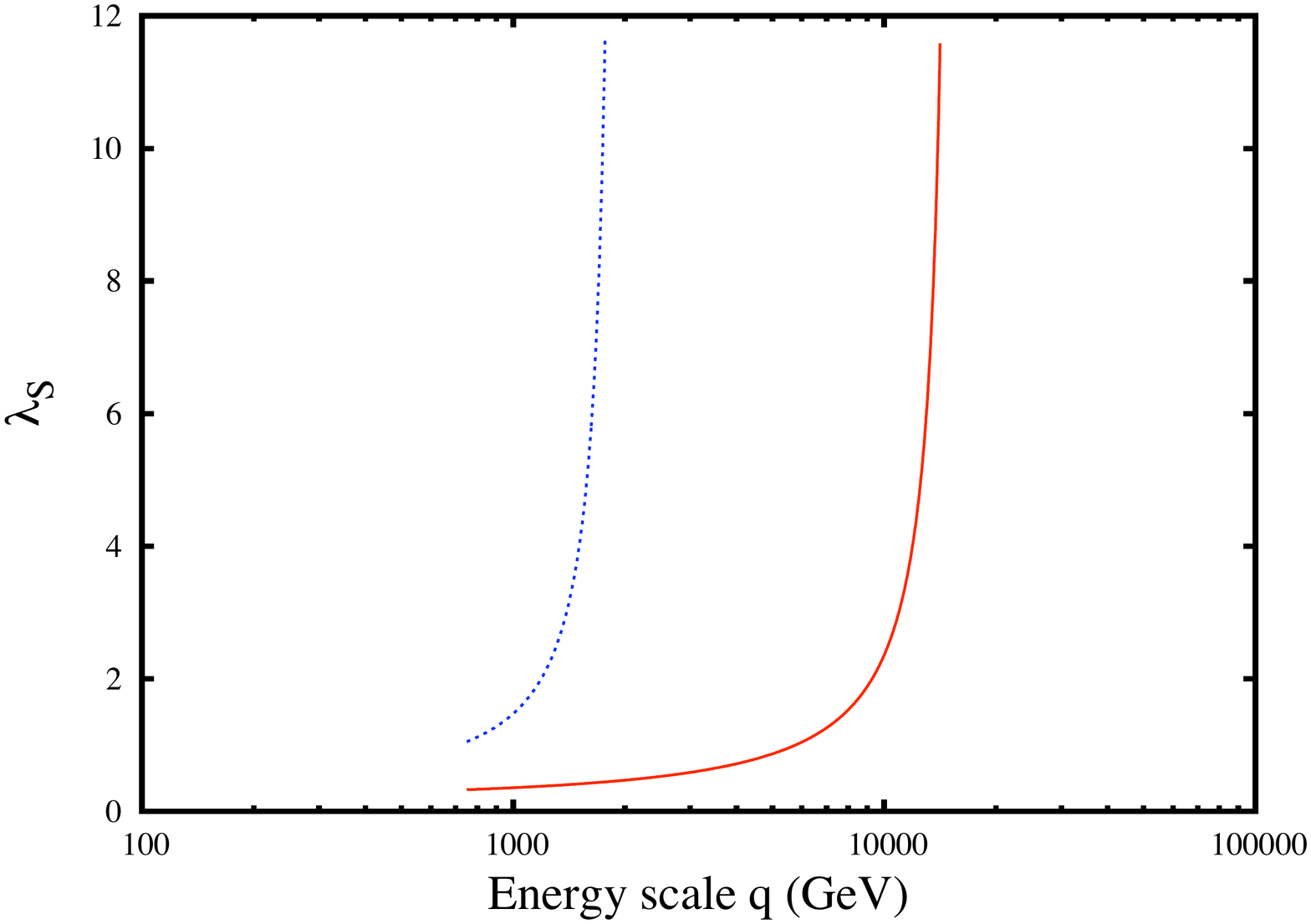}
\includegraphics[width=5.5cm]{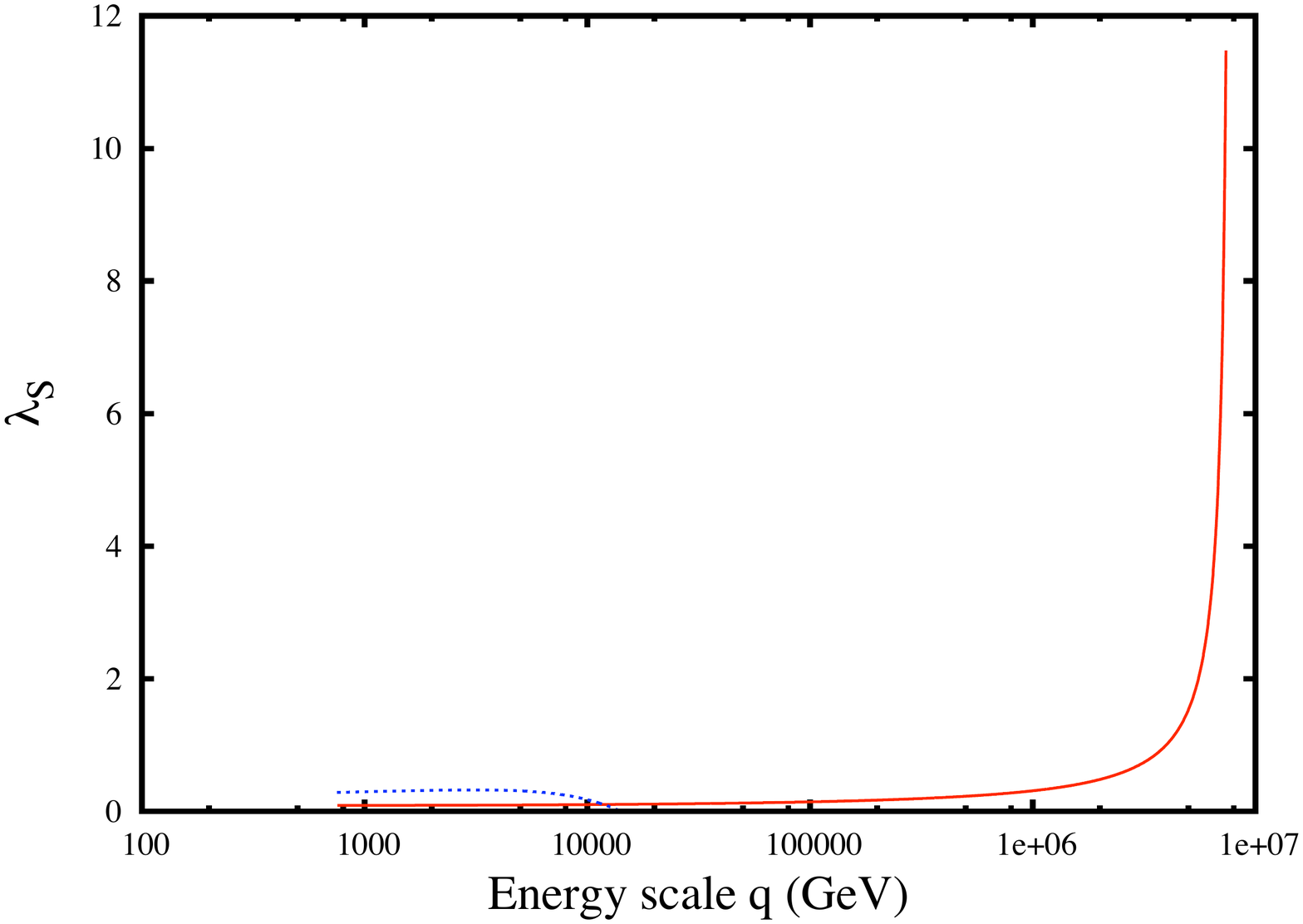}
\includegraphics[width=5.5cm]{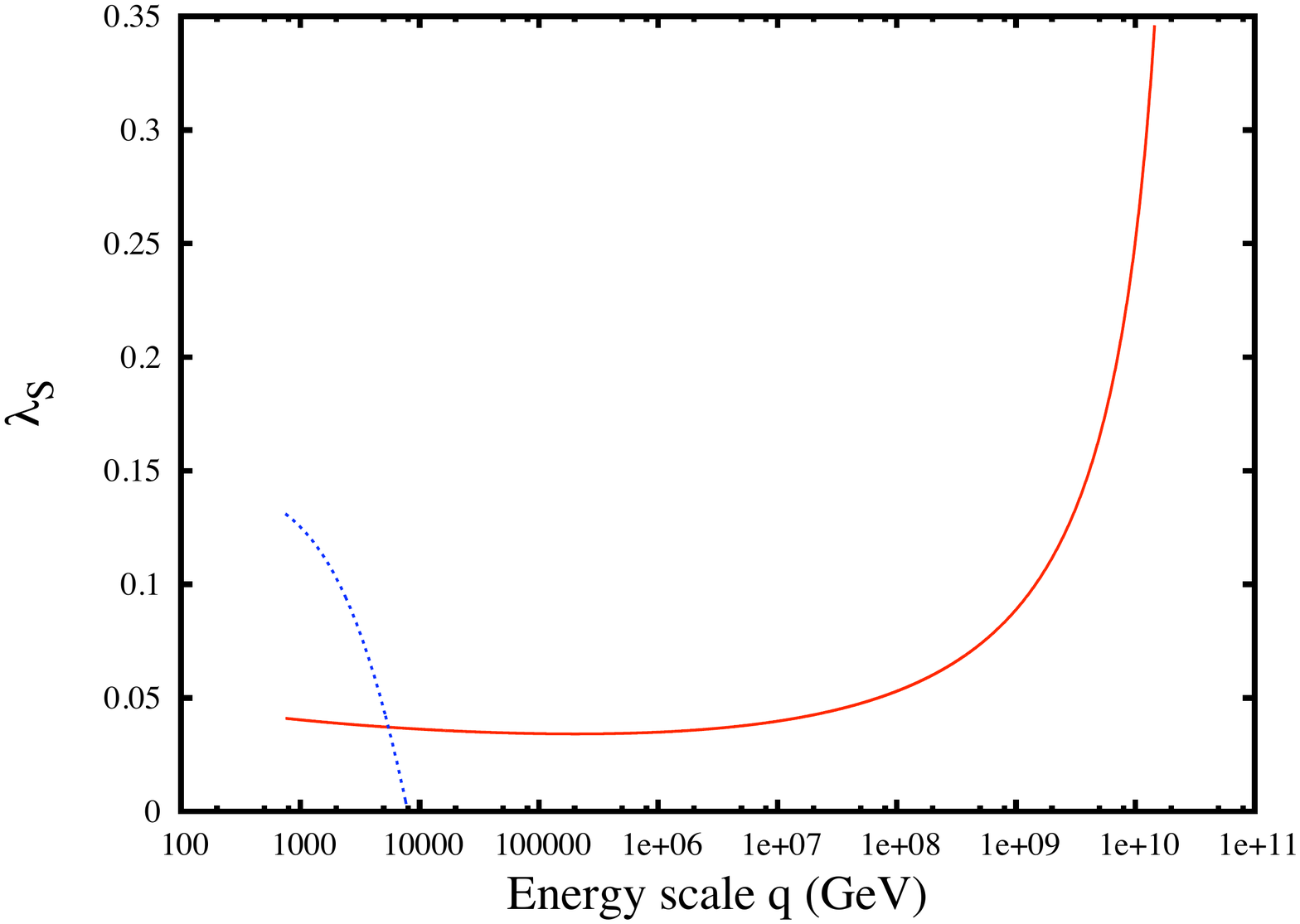}
\end{center}
\caption{\em\small The running of $\lambda_S$ with energy as a function of $N$, for $N=10$ (left), 20 (centre) and 30 
(right). The solid (red) line is for the lower range of the diphoton cross-section, and the dashed (blue) line is for the 
upper range. We have taken only one
charge $+\frac23$ isosinglet vector-like quark at $M_Q=1$ TeV.}
\label{fig:rg}
\end{figure}


Interesting but expected results appear when we look at the renormalization group evolution of the couplings, 
as shown in Fig.\ \ref{fig:rg}, for 
which we use the one-loop $\beta$-functions given in Eq.\ (\ref{all-rge}). We use the following values for the 
scalar quartic couplings, $\lambda$ being already fixed by the Higgs boson mass.
\bea
N=10 &\Rightarrow& a = 0.42\,,\ \ \lambda_S = 0.33 (1.05)\,,\nonumber\\
N=20 &\Rightarrow& a = 0.21\,,\ \ \lambda_S = 0.09 (0.29)\,,\nonumber\\
N=30 &\Rightarrow& a = 0.14\,,\ \ \lambda_S = 0.04 (0.13)\,.
\eea
These numbers might be taken as some sample benchmark values, the lower (higher) value of $\lambda_S$ being 
chosen for the lower (higher) range of the diphoton cross-section, $\sim 3$ fb ($\sim 9$ fb). 
The value of $\zeta_Q$ is fixed by 
$N$, $e_Q$, the cross-section, and the mass of $Q$, which we take to be 1 TeV. 
However, one can also check that these benchmark values make the coefficients of the quadratic divergences 
of the scalars to vanish at the energy scale of 750 GeV, whatever might be read into that.

Obviously, higher 
values of the couplings make the model unstable more quickly. For $N=10$, the upper value of 
$\lambda_S$ hits the Landau pole at 1.8 TeV, which is very much
accessible at the LHC; for the lower value, the range is a bit higher, at about 14 TeV. Thus, even with 10 such singlets, the model 
is superseded by some ultraviolet complete theory at a few TeV at the most, and the situation is much worse for 
smaller values of $N$. One may postpone this fate till a higher energy scale with smaller values of $a$ to start with.  
On the other hand, smaller values of $\lambda_S$ will make the theory unstable even faster, because of the 
negative pull of the Yukawa coupling $\zeta_Q$. 

The problem is easy to identify: the large value of $\zeta_Q$ needed to satisfy the data. One may make the model 
work till the Planck scale with enough singlet scalars (the exact number depends on the values of $a$ and $\lambda_S$),
also leading to a first-order electroweak phase transition. However, the situation improves considerably if 
the diphoton cross-section settles down to the lower range,
something like the average of 8 and 13 TeV data.

\section{Summary}

In this paper, we have tried to take a critical look at one of the simplest models to explain the recently 
observed diphoton excess at the LHC, namely, a singlet scalar associated with a vectorlike fermion. The immediate 
hurdle is the large number of events, which means a large production cross-section, and hence a Yukawa 
coupling $\zeta_Q$ which 
is nonperturbative. This also means that a renormalizable theory has very limited validity, thus indicating the possible
presence of more scalars and fermions.

With one singlet scalar, both $\zeta_Q$ and $\lambda_S$ are strongly nonperturbative. While one cannot apply the 
perturbative $\beta$-functions to evaluate the running of the couplings, one may surmise that the model is quite 
unstable because of the strong negative pull of $\zeta_Q$ on $\lambda_S$. This forces us to consider a scenario where
there are more than one such singlet scalars. They better be quasi-degenerate, all contributing to the signal, 
and explaining the broad resonance as a sum of unresolved narrow resonances. In the simplest scenario, the singlets do 
not mix with the SM doublet field. This removes the constraint coming from $t\bar{t}$ production at the resonance, 
and keeps the Higgs boson partial decay widths consistent with the SM expectation. 

With $N \gg 1$, the couplings are brought back in the perturbative region, and a study of the renormalization group 
equations indicate that the model remains valid beyond the LHC range, the validity depending strongly on $N$ and 
rapidly increasing with it. 

One may fine-tune or extend the model further, by adding more vectorlike fermions. Even the simplest model with 
one isosinglet quark is not realistic without a tiny admixture with the chiral quarks. One may have 
$SU(2)$ doublets or triplets 
of such vectorlike quarks, and also vectorlike leptons.  
If a dijet excess is not seen at 750 GeV,
vectorlike leptons with stronger Yukawa couplings than the quarks may be a solution. There may be more than one 
generation of such fermions. The $O(N)$ symmetry of the scalars may be softly broken. The next step would be to 
confirm this signal, look for the dijet excess, followed by a search for exotic fermions. Unfortunately, if the 
singlet scalars are quasi-degenerate, the resolution of LHC may not be enough to split the individual peaks. 
Another interesting possibility is to study the model in a photon-photon collider \cite{phph}.

\centerline{\bf{Acknowledgements}}

I.C.\ acknowledges the Council for Scientific and Industrial Research, Government of India, for a 
research fellowship. A.K.\ acknowledges the Department of Science and Technology, Government of  
India, and the Council for Scientific and Industrial Research, Government of India, for support 
through research grants.

\end{document}